\RequirePackage{fix-cm}
\documentclass[natbib,twocolumn]{svjour3}                     
\smartqed 

\journalname{Hydrobiologia} 

\usepackage{overpic}
\usepackage{paralist}

\newcommand{\species}[1]{\textit{#1}}
\newcommand{\mytilus}{\species{Mytilus edulis}}
\newcommand{\edulis}{\species{M.\ edulis}}

\newcommand{\science}[2]{\ensuremath{#1\cdot10^{#2}}}
\newcommand{\degree}{\ensuremath{^{\circ}}}

\newcommand{\reftab}[1]{Table~\ref{tab:#1}}
\newcommand{\refsec}[1]{Sect.~\ref{sec:#1}}
\newcommand{\reffig}[1]{Fig.~\ref{fig:#1}}

\newcommand\mpcbmd{\ensuremath{\textrm{mmol\,C\,m}^{-3}\,\textrm{d}^{-1}}}
\newcommand\mpcbm{\ensuremath{\textrm{mmol\,m}^{-3}}}

\newcommand\gpy{\ensuremath{\textrm{g\,m}^{-2}\,\textrm{a}^{-1}}}

\newcommand\pcbm{\ensuremath{\textrm{\,m}^{-3}}}
\newcommand\TPM{\ensuremath{\textrm{TPM}}}

\newcommand\CR{\ensuremath{\textrm{CR}}}
\newcommand\FR{\ensuremath{\textrm{FR}}}

\usepackage{hyperref}

\usepackage[nostamp]{draftwatermark}
\SetWatermarkAngle{90.0}
\SetWatermarkLightness{0.0}
\SetWatermarkFontSize{14pt}
\SetWatermarkHorCenter{2.4\hsize}
\SetWatermarkText{Revised manuscript subm. to Hydrobiologia 9 May 2018.  Do not cite without consulting  authors}
\SetWatermarkColor{red}
\begin{document}

\title{The large scale impact of offshore wind farm structures on pelagic primary productivity in the southern North Sea
\thanks{
This research is funded by the Marine, Coastal and Polar Systems (PACES~II) of the Hermann von
Helmholtz-Gemeinschaft Deutscher Forschungszentren e.V..  K.S. is funded by the European Commission Erasmus Mundus Masters Course in Environmental Sciences, Policy and Management (MESPOM).  C.L., O.K. and K.K. received support from the ``Modular System for Shelves and Coasts'' (MOSSCO) grant provided by the Bundesministerium f\"ur Bildung und Forschung under agreements 03F0667A and 03F0667B;  O.K. and K.W. are also supported by the DFG priority programme~1704 ``Flexibility matters: Interplay between trait diversity and ecological dynamics using aquatic communities as model system'' (DynaTrait) under grant agreement KE~1970/1-1. K.K. is furthermore supported by the DFG Collaborative Research Center ``Energy Transfers in Atmosphere and Ocean'' TRR181. We thank all co-developers of the model coupling framework MOSSCO, foremost M. Hassan
Nasermoaddeli and Richard Hofmeister.  The authors gratefully acknowledge the computing time
granted by the John von Neumann Institute for Computing (NIC) and provided on the supercomputer
JURECA at Forschungszentrum J\"ulich.  We are grateful to the open source community that
provided many of the tools used in this study, including but not limited to the communities developing
ESMF, FABM, GETM and GOTM.
}}

\titlerunning{Offshore wind farm impact on North Sea primary productivity}

\author{
Kaela Slavik
\and Carsten Lemmen
\and Wenyan Zhang
\and Onur Kerimoglu
\and Knut Klingbeil
\and Kai W. Wirtz
}

\institute{K. Slavik \at Helmholtz Zentrum Geesthacht Zentrum f\"ur Material- und K\"ustenforschung, Germany \\  
\emph{Present address: Future Earth Paris Global Hub, Universit\'e Pierre et Marie Curie, France} 
\and C. Lemmen \at Helmholtz Zentrum Geesthacht Zentrum f\"ur Material- und K\"ustenforschung, Germany\\
\emph{Corresponding author} \\
Tel.: +49 4152 87-2013\\ Fax: +49 4152 87-2020\\ \email{carsten.lemmen@hzg.de} 
\and W. Zhang, O. Kerimoglu, K.W. Wirtz \at  Helmholtz Zentrum Geesthacht Zentrum f\"ur Material- und K\"ustenforschung, Germany
\and K. Klingbeil \at  Universit\"at Hamburg, Germany and Leibniz-Institut f\"ur Ostseeforschung Warnem\"unde, Germany
}

\date{Received: date / Accepted: date}

\authorrunning{Slavik et al.}

\maketitle

\begin{abstract}

The increasing demand for renewable energy is projected to result in a 40-fold increase in offshore wind electricity in the European Union by~2030. Despite a great number of local impact studies for selected marine populations, the regional ecosystem impacts of offshore wind farm structures are not yet well assessed nor understood. Our study investigates whether the accumulation of epifauna, dominated by the filter feeder \textit{Mytilus edulis} (blue mussel), on turbine structures affects pelagic primary productivity and ecosystem functioning in the southern North Sea. We estimate the anthropogenically increased potential distribution based on the current projections of turbine locations and reported patterns of \edulis{} settlement. This distribution is integrated through the Modular Coupling System for Shelves and Coasts to state-of-the-art hydrodynamic and ecosystem models. Our simulations reveal non-negligible potential changes in regional annual primary productivity of up to $8\%$ within the offshore wind farm area, and induced maximal increases of the same magnitude in daily productivity also far from the wind farms. Our setup and modular coupling are effective tools for system scale studies of other environmental changes arising from large-scale offshore wind-farming such as ocean physics and distributions of pelagic top predators.

\end{abstract}

\section{Introduction}
\label{sec:intro}

Recognition of the role of burning fossil fuels in anthropogenic climate change has led to 
increased investment in renewable energy such as  wind farming \citep{Edenhofer2011}. In particular,
offshore wind energy has proliferated over the past decade and will be integral in the transition
to renewable energy systems. In the European Union (EU), offshore wind farms (OWFs) are predicted to increase
13-fold between 2015 and 2020, and 40-fold by~2030, in order to meet 4.2\% of EU total electricity
consumption \citep{GWEC2015}.  

\begin{figure}
\centering
\begin{overpic}[viewport=0 105 420 405,clip=,width=.75\textwidth]{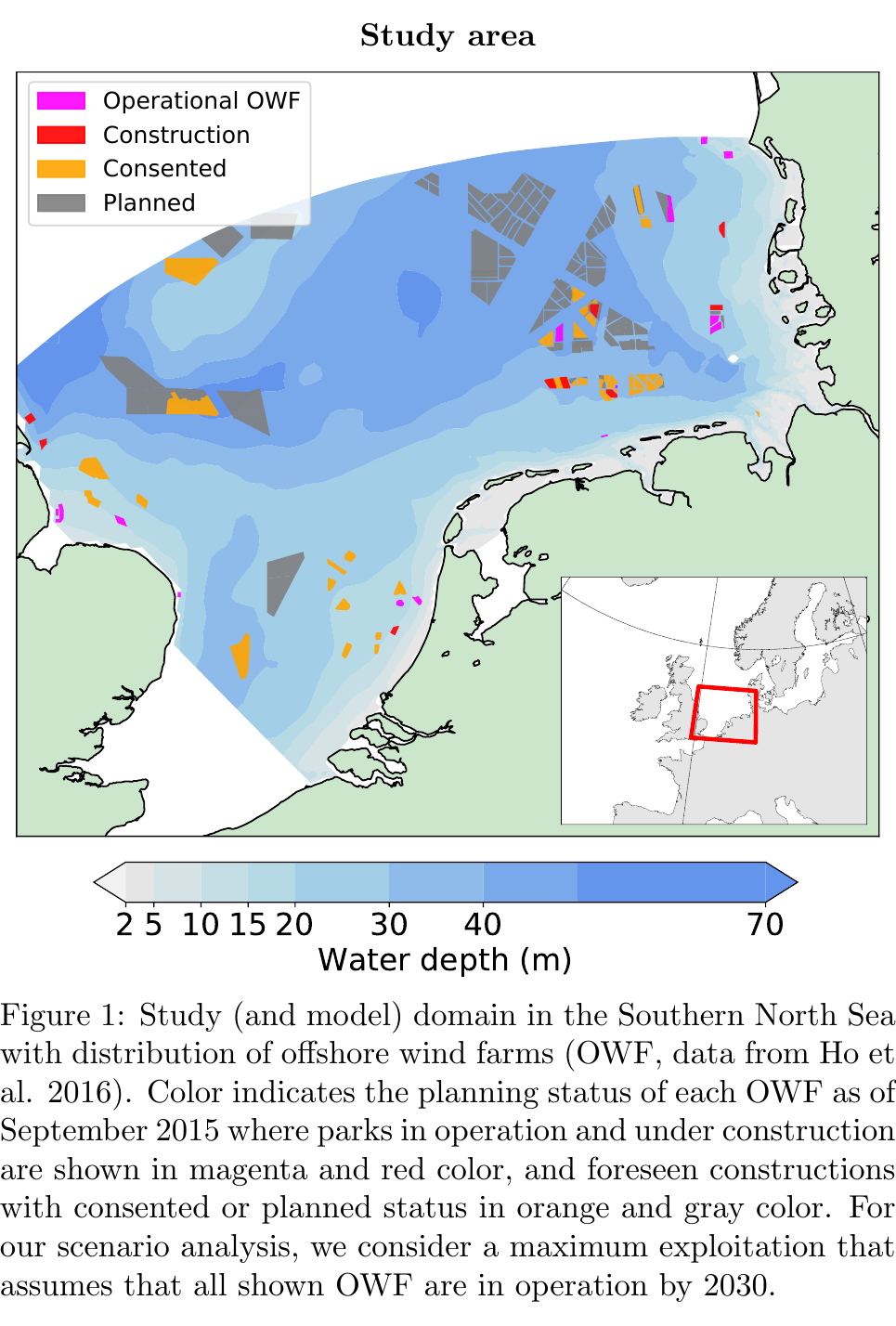}
\end{overpic}
\caption{Study (and model) domain in the Southern North Sea with distribution of offshore
wind farms (OWFs, data from \citealt{EWEA2016}).  Color indicates the planning status of
each OWF as of September 2015
where parks in operation and under construction are shown in magenta and red color, and
foreseen constructions with consented or planned status in orange and gray color.  For our scenario
analysis, we consider a maximum exploitation that assumes that all shown OWFs are in operation
by 2030.}
\label{fig:statusdistribution}
\end{figure}

Currently, ~63\% of OWFs in EU waters are concentrated in the southern North Sea (SNS),
with the  remainder located in the Atlantic Ocean and Baltic Sea. The SNS is expected to
remain a hotspot for EU OWF development, where $\approx 85\%$ of OWFs are
currently under construction and $\approx 75\%$ of OWFs have received consent \citep{EWEA2016}.
Offshore wind turbines are solid structures, typically larger than their onshore counterparts,
built of steel or concrete, with monopiles accounting for 80\%.
OWFs are being built further offshore and in deeper waters, with the average water depth
increasing three-fold and the average distance from shore five-fold between 1991 and
2010 \citep{Kaldellis2013}. The large additional build-up of OWFs by 2030 is 
evident from the spatial distribution of operational, under construction, consented and
planned OWFs in the SNS (\reffig{statusdistribution}).

\begin{table*}
\centering
\small
\caption{Offshore wind farms in the Southern North Sea where \edulis{} are the dominant species.}
\label{tab:dominantspecies}
\begin{tabular}{lll}
\hline\noalign{\smallskip}
Country & Location & Reference  \\
\noalign{\smallskip}\hline\noalign{\smallskip}
Germany    & FINO 1 research platform & \citealt{Krone2013}\\
Belgium      & C-Power OWF & \citealt{Kerckhof2012}\\
Netherlands & Egmond aan Zee OWF & \citealt{Bouma2012,Lindeboom2011}\\
Denmark	 & Horns Rev OWF & \citealt{Leonhard2006}\\
Sweden 	 & West coast of Sweden & \citealt{Langhamer2009}\\
\noalign{\smallskip}\hline
\end{tabular}
\end{table*}

The increasing number of OWFs alters the functioning of the 
surrounding pelagic ecosystem by restructuring the biological communities at and
around the submerged foundations and pile structures \citep{Joschko2008,Krone2013}. Specifically,
they increase the biomass and distribution of filter feeders
\citep{Krone2012,Lindeboom2011}, because OWFs provide the hard substrate needed
for colonisation by a variety of epistructural species. This colonisation is also referred to as biofouling. 
Among the colonisers,  the bivalve \textit{Mytilus edulis} (L.~1758, Bivalvia: Mytilidae)
is the dominant species near the water surface (\reftab{dominantspecies})
\citep{Freire1995,Riis2003,Wilhelmsson2008,Joschko2008,Krone2013}. For example, piles of the OWF research platform 
FINO~1 (Forschungsplattformen in Nord- und Ostsee) are
covered by an average of 4300\,kg of \edulis{}, with a turnover rate of more
than 50\% of the stock per year \citep{Krone2013}. 

Biofouling not only generates habitat for a new epistructural community, i.e. biota that live on and attach to
a structure, but it has further
consequences for the benthic community underneath and the surrounding pelagic zone
\citep{Krone2012,Maar2009}. Filter feeders have been
shown to significantly reduce the ambient 
concentration of phytoplankton and  of micro- and mesozooplankton
\citep{Dolmer2000,Maar2007}, which to some extent likely applies to epistructural  \edulis{} as well
 \citep{Maar2009}.  By changing
phytoplankton biomass, epistructural filtration can be expected to affect primary productivity and thus the very basis of
the marine food web and biogeochemical cycling locally above mussel beds and around
the offshore wind turbine.

Our study aims to assess the sensitivity of pelagic primary productivity to changed abundance and distribution of
\edulis{} on OWFs for an entire regional-scale ecosystem.  It is the first study
to investigate the accumulated effects on primary productivity at the systems scale, beyond the local impacts of
individual offshore wind turbines.
Prerequisites for such an assessment are \begin{inparaenum}[(i)]
\item the reconstruction of\edulis{} abundance both for their natural, epibenthic habitat and for the new
epistructural niches;
\item the functional coupling of the lateral and vertical distribution of reconstructed mussels 
to phytoplankton prey fields in a realistic hydrodynamic and biogeochemical representation of the SNS.  
\end{inparaenum}

For the integrated modelling of benthic and epistructural filtration, water physics and pelagic
biogeochemistry, we use the recently introduced modular framework by 
\citet{Lemmen2018}, which contains a novel ecosystem model recently applied to and verified for the SNS by 
\citet{Kerimoglu2017}.  Multi-annual simulations run with and without epistructural mussels allow
a first estimate of the sensitivity of pelagic primary productivity to the projected OWFs
in this regional sea.

\section{Materials and Methods}\label{sec:materials}

\subsection{Study location}\label{sec:location}
The southern North Sea (SNS) is located between 51\degree\,N and 56\degree\,N
and is bordered by the United Kingdom, Belgium, the Netherlands, Germany and Denmark (\reffig{statusdistribution}).
The water is fairly shallow  with an average depth of $30$\,m and comprises
an extended area of intertidal flats and several major estuaries \citep{Eisma1987}.  
The seabed is  composed predominantly of sand and, in the deeper and more coastal
parts, of mud \citep{Walday2002}.  The SNS experiences strong seasonal variability, with
winter storms often generating large surface waves and suspending greater amounts of
sediments \citep{Groll2017,Nasermoaddeli2017}.  Currents in the North Sea are generated by tides and wind forcing,
with the latter especially important during storm events  \citep{Howarth2001}.  
The North Sea obeys a general cyclonic circulation. This is driven by prevailing westerly winds, 
residual tidal currents and the baroclinic pressure gradient set up by coastal river discharge \citep[e.g.,][]{Otto1990}. 
The residual circulation
within the basin flows southward along the east coast of the UK, before turning west in the East Anglia plume
and then continuing westward along the West Frisian barrier islands. Part of the residual current then
continues northward towards Norway.  The other part continues along the East Frisian barrier islands and
joins the Elbe and Weser River inflows.  It then turns northwest again towards the central North Sea,
bypassing Helgoland Island, before turning back towards and flowing north along the Danish coastline \citep{Carpenter2016}.  

\subsection{Reconstruction of spatial distribution of epibenthic \edulis{}}\label{sec:reconstructed}
  
Open access spatial data on the abundance and distribution of \edulis{} were obtained from
the Joint Nature Conservation Committee (JNCC), the Ocean Biogeographic
Information System (OBIS), the Archive for Marine Species and Habitats Data
(DASSH), the Global Biodiversity Information Facility (GBIF) and the Belgian
Marine Data Centre (BMDB).  Most of the data (43\%, rounded) was from JNCC, 26\% and
23\% from BMDC and GBIF (containing presence only data), and 9\%
from OBIS.  Only few  data points came from DASSH $<1$\%.  Removing duplicate
locations, in total 4074 count observations and 37\,214 presence only data
were used for the reconstruction of the spatial distribution of epibenthic \edulis{}.

To extrapolate and interpolate the count and occurrence data to the entire domain
of the SNS, we used empirical relationships between mussel abundance, sediment
grain size and depth. We added to this a low abundance random distribution for deep water
and a constant high abundance for mussel beds.
As \edulis{} are tolerant to large variations in temperature ($0$--$29$\,\degree C)
and salinity \citep{Seed1992}, such factors were not considered in the reconstruction.
Taking the average adult \edulis{} individual biomass as $600$\,mg dry weight (DW)
\citep{Bayne1980}, which equals $64.5$\,mg ash-free dry weight 
\citep[AFDW, ][Table 2]{Ricciardi1998}, the abundance and distribution of \edulis{} in the SNS was spatially
reconstructed using the median sediment grain size map that is publicly available
from the NOAH habitat atlas (\href{http://www.noah-project.de/habitatatlas/}{www.noah-project.de/habitatatlas/}). 

\edulis{} prefers larger sediment grain sizes and hard substrate \citep{OSPAR2010}, 
thus an increase in abundance density ($n$) with increasing sediment grain size, ranging from
an abundance of $1$\,m$^{-2}$ in muddy areas (median grain size $d_{50} < 
0.06$\,mm) to $40$\,m$^{-2}$ in areas of coarse gravel, at locations where mussels are found. We employed
a Random Forest model \citep{Liaw2002} to create a predictor of abundance density from median grain size.
Comparison to Wadden Sea field data compiled by \citet{Compton2013}, however, indicated that
predicted shallow-water mussel abundance was greatly overestimated, which can be attributed
to a positive sampling bias in the citizen-science data set.  We thus interpreted
the count data as relative, i.e. as a probability of occurrence that needs to be rescaled to conform to the 
\citet{Compton2013} estimate where it borders the Wadden Sea, and rescaled the data accordingly.

The abundance--sediment
grain size relationship is applied up to a $10$\,m natural depth limitation
\citep{Reise1987,Suchanek1978}. Outside the depth limitation, \edulis{} still occur,
however at a much reduced density, and are often completely absent:  A random
density between $0$\,m$^{-2}$ and $0.5$\,m$^{-2}$ is assigned.   In the Wadden Sea, no sediment 
data is available in the NOAH data set, and a constant value of $2$\,m$^{-2}$ is assigned on 
the Wadden flats consistent with  \citet{Compton2013}.  Mussel beds
were incorporated as point data using the OSPAR Biodiversity Committee habitat classification,
where a constant density of $3911$\,m$^{-2}$ \citep{Nielsen2007} is  downscaled 
to $170$\,m$^{-2}$  to account for the patchiness of the beds.  

Presence only data is not a preferred estimator for species distribution modelling, especially when there is
a sampling bias.  Many of the GBIF-reported \edulis{} observations are opportunistic finds reported by
citizen scientist divers, with a bias towards more accessible near-coast areas and towards summer temperature.  
This bias may
be overcome by environmental constraints that can serve as proximate absence \citep{Phillips2009}, such
as water depth for \edulis{}.  We note that the epibenthic reconstruction of abundance
presented here is preliminary.  As it serves as a baseline only, the uncertainty in this epibenthic reconstruction
does not harm the results obtained for the ecosystem sensitivity (see \refsec{dominantspecies}).  
We are currently working on a refined epibenthic reconstruction that address the effect of this uncertainty
on the baseline itself (Lemmen, North Sea ecosystem-scale quantification of primary productivity changes
by the benthic filter feeder \mytilus, unpublished manuscript).

\subsection{Epistructural \edulis{}}\label{sec:dominantspecies}

The biomass and species diversity of epistructrual communities at OWFs are much
higher than would be found on natural hard substrate  \citep{Wilson2009}, with species
composition varying with both depth and time, as recorded at both FINO~1 
\citep{Krone2013,Joschko2008}, and the Kentish Flats OWF  \citep{Bessel2008}.  \edulis{} is 
the dominant macrofauna species at shallower depths, while at greater depths Anthozoa and
\species{Jassa spp.}  are more prolific;   other major taxa such
as green algae, \species{Asterias rubens} (Asteroidea), Bryozoa, Porifera and
\species{Tubularia spp.} are also present \citep{Krone2013}.  \edulis{} is the most abundant and
ecologically important species at OWF epistructrual communities in the North Sea
(\reftab{dominantspecies} and \citealt{Borthagaray2007}), contributing up to 90\% of
epistructural biomass in some locations. It is therefore also the main
driver of ecological change around offshore structures \citep{Krone2013,Maar2009}.  

\begin{table}
\small\centering
\caption{\edulis{} biomass with depth, averaged over all years 2005--2007 sampled by  \citet[][, pp. 4--5]{Krone2013}.}
\label{tab:pile}
\begin{tabular}{lp{2cm}p{2cm}p{2cm}}
\hline\noalign{\smallskip}
Depth (m) & Distribution (\%) & range biomass density (kg m$^{-2}$)& mean biomass total (kg) \\
\noalign{\smallskip}\hline\noalign{\smallskip}
0.0 -- 2.5	& 95	& 22.3--43 & 3258.08 \\
2.5 -- 7.5	& 3	& 0.5--3.9 & 58.58 \\
7.5 -- 15	& 2	& n/a & 19.29 \\
15.0 -- 30.0	& n/a & 0	& 1.63 \\
\noalign{\smallskip}\hline
\end{tabular}
\end{table}

The additional settlement of \edulis{} as a result of OWFs is considered by
incorporating the vertical distribution observed by \citet{Krone2013} at the FINO~1
OWF.  \edulis{} abundance ($n$) at an offshore wind turbine is a function of its
radius ($r$) and its base depth ($z$), with the radius assumed to be $3$\,m at all
OWFs \citep{Orbis2013}.  The influence of \edulis{} on water properties is
assumed to be equal around the entire circumference, without consideration of current
direction.  Multiplying the abundance density by the circumference gives the vertical
distribution of \edulis{} with depth at offshore wind turbines (\reftab{pile}). 
The abundance density over depth at each offshore wind turbine was calculated by
converting the wet weight reported by \citet{Krone2013} to DW using a factor
of $6.6\%$ and assuming $600$\,mg\,DW ind$^{-1}$ \citep{Ricciardi1998,Bayne1980}.  
We did not consider annual variation despite observed seasonal variations in the data 
set by  \citet{Krone2013} because mussel biomass sampled at different seasons over the 
years 2005--2007  were not found to be significantly different.

\subsection{Spatial subgrid distribution}

\begin{figure*}
\begin{overpic}[viewport=0 100 430 275,clip=,width=\textwidth]{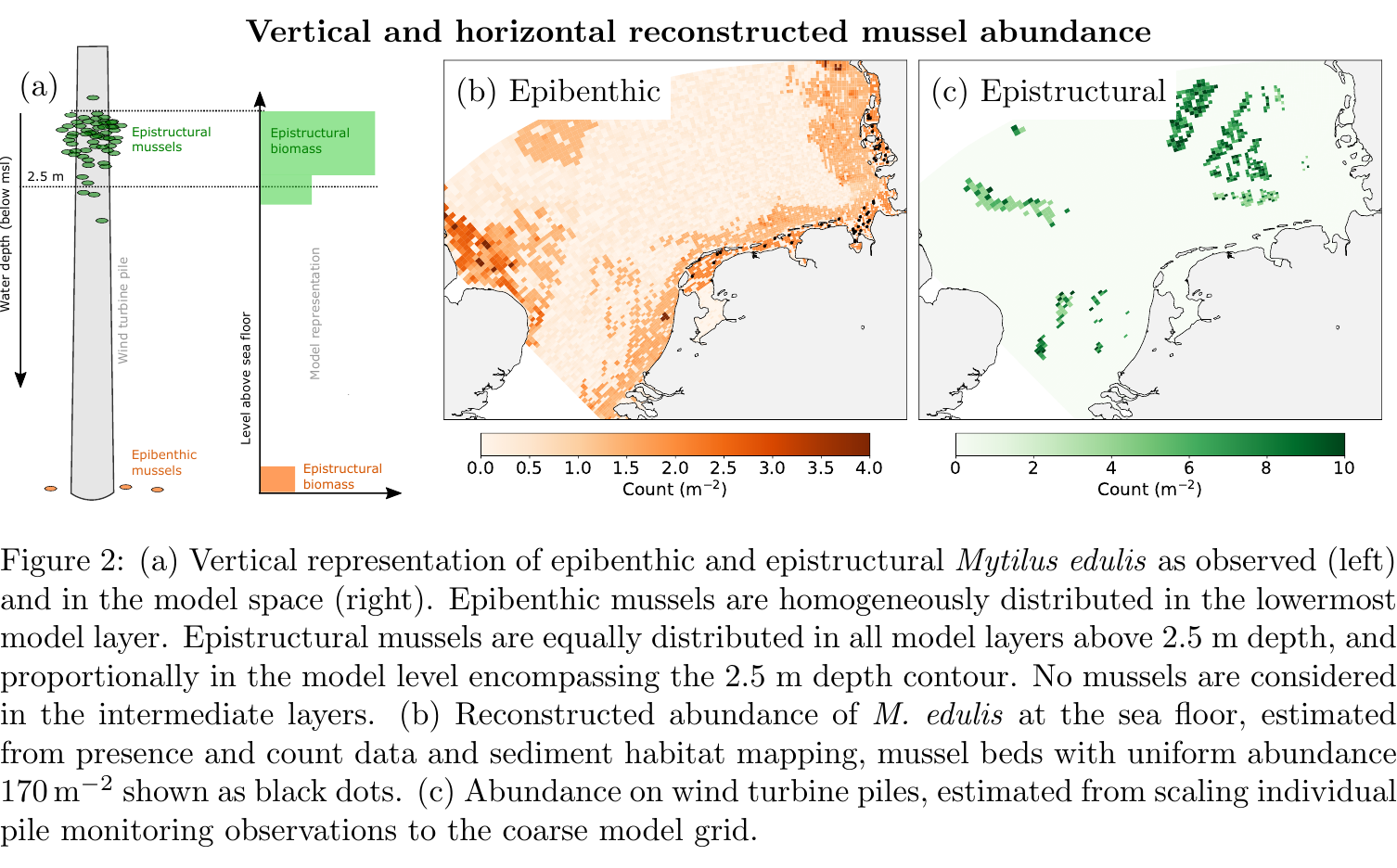}
\end{overpic}
\caption{(a) Vertical representation of epibenthic and epistructural \edulis{} as observed (left) and in the model space (right).  Epibenthic mussels are homogeneously
distributed in the lowermost model
layer. Epistructural mussels are equally distributed in all model layers above 2.5 m depth, and
proportionally in the model level encompassing the 2.5 m depth contour.  No
mussels are considered in the intermediate layers. (b) Reconstructed abundance of \edulis{} at the sea floor, estimated from presence and count data and sediment
habitat mapping; mussel beds with uniform abundance 170\,m$^{-2}$ shown as black dots. (c) Abundance on wind turbine piles, estimated from scaling individual pile monitoring observations to the coarse model grid.}
\label{fig:musseldistribution}
\end{figure*}

The spatial distribution of current and projected OWFs in the southern North Sea
(\reffig{statusdistribution}) was overlaid on a curvilinear grid later used for the 
numerical model.  Epibenthic areal abundance of \edulis{} was considered to be (vertically) 
equally distributed within the lowermost representable physical layer in the grid of
the hydrodynamic model.   Epistructural areal abundance was vertically distributed
in all simulation layers representing the upper $2.5$\,m of the water column in the hydrodynamic model
(\reffig{musseldistribution}a). 

Estimating abundance density at OWFs from the abundance at individual turbines
requires consideration of the turbine density at OWFs. Offshore wind turbines are
often spaced between five and eight times the rotor diameter \citep{Eon2011}, which
tend to range from $80$ to $100$\,m \citep{IRENA2012}.  Assuming a rotor diameter
of $100$\,m and a spacing of five times this distance, this means that each offshore
wind turbine requires $500$\,m spacing or $0.25$\,km$^{2}$ area, giving an average
wind turbine density of $4$\,km$^{-2}$.  

\subsection{Coupled model system}
Extrapolation from the compiled OWF locations to the entire SNS area and the
description of \edulis{} influence on the pelagic ecosystem requires a
spatially explicit, coupled model approach, for which we employ the recently
developed open source software infrastructure Modular System for Shelves
and Coasts (MOSSCO, \href{http://www.mossco.de}{www.mossco.de}, \citealt{Lemmen2018}).
MOSSCO facilitates the exchangeable coupling of models and data sets
and enables the integration of modules describing
physical, chemical, geological, ecological and biogeochemical processes. 
MOSSCO applications for the 3D coastal ocean focus on processes at the
benthic--pelagic interface and, among others, explain spatio-temporal patterns in
coastal nutrient concentration \citep{Hofmeister2017,Kerimoglu2017}, primary
productivity \citep{Kerimoglu2017}, macrobenthic biomass and community dynamics
\citep{Zhang2017} and suspended
sediment concentration as affected by macrobenthic activities \citep{Nasermoaddeli2017}.

MOSSCO features generic output and input
components that can be used to integrate, e.g., river nutrient fluxes, open ocean boundary
conditions and  faunal abundance.  As a physical
driver within MOSSCO, we employed the coastal ocean model GETM (General Estuarine Transport Model,
\citealt{Burchard2002jsr,Klingbeil2013}) to calculate sea level, currents, temperature and
salinity distributions, and to transport the biogeochemical and ecological quantities.
GETM obtains state-of-the-art turbulence closure from the General Ocean Turbulence
Model \citep[GOTM,][]{Umlauf2005}, and has been shown to have  high skill in various studies
for the North Sea and SNS 
\citep[e.g.][]{Graewe2016,Purkiani2016}.

Pelagic ecology was described by the Model for Adaptive Ecosystems in Coastal
Seas (MAECS, \citealt{Wirtz2016}) implemented as a  FABM module;
MAECS simulates pelagic nutrient, phytoplankton, zooplankton and detritus dynamics 
and accounts for the acclimation of  intracellular composition in phytoplankton.
In our application, MAECS resolves the elements carbon (C), nitrogen (N), and phosphorus (P), and features adaptive shifts in phytoplankton ecophysiology as described by, e.g., variable chlorophyll a (Chl-a) and RuBisCO contents. The underlying scheme for these adaptive shifts has been derived as an optimality theory and was first applied to phytoplankton growth and succession by \cite{Wirtz1996}. Pelagic element fluxes are described similar to other ecosystem models including nutrient uptake during phytoplankton growth, transformation through phytoplankton mortality including herbivorous grazing, and stoichiometrically controlled turnover of detritus and dissolved organic matter in terms of C, N, and P.

 A full description and an extensive performance assessment of the model for a decadal hindcast of
the SNS has been provided by \citet{Kerimoglu2017}. Our coupled setup differs in two respects:
(i) we resolve  filtration (see \refsec{filtrationmodel}), and (ii) we used the full 3D biogeochemical 
model OmexDia based on \citet{Soetaert1996a} instead of the the single layer soil parameterization by 
\citet{Kerimoglu2017}.   There, top-down mortality of zooplankton is uniform, while we prescribe
higher zooplankton mortality near the coast.  Furthermore, the ecosystem model MAECS has since evolved and now includes a parameterization for viral loss of phytoplankton (Wirtz 2018, Physics or biology? Persistent chlorophyll accumulations in a shallow coastal sea explained by pathogens and carnivorous grazing, submitted manuscript, hereinafter referred to as Wirtz, submitted).

\subsubsection{Filtration model}\label{sec:filtrationmodel}
\edulis{} actively passes water over a specialized filtering structure (the gill), thereby
removing a significant proportion of both organic (i.e., mainly phytoplankton) and inorganic
particles with high efficiency \citep{Widdows1979}. The volume of water passed
over the gill area per unit of time and individual body volume is referred to as the
clearance rate (\CR). CR has been observed to increase with rising current velocity 
\citep{Cranford1999}.  At very low ambient Chl-a concentration below about
0.5\,mg\pcbm{}, however, CR ceases for energetic reasons \citep{Riisgard2003}.
The removal of particles from the cleared water, termed the filtration rate (\FR),
depends, among others, on the concentration and organic quality of
particles.  A physiological regulation of filtration rate is, however, debated 
 and has been
studied for high ambient food concentrations only. At the concentrations typically found in the
SNS, full exploitation of the ambient concentration can be expected \citep{Clausen1996,Asmus1991}.

Our model implementation of \edulis{} FR is based on the empirical relations identified by
\citet{Bayne1993}.  They formulated the relationship in terms
of phytoplankton carbon amount concentration ($[C]$) and total particulate matter (\TPM)
relative to an assumed individual DW of 300\,mg. 
\begin{equation}
\FR_{{\small\TPM},300} = 0.05 \cdot[C]^{0.983}, 
\end{equation}
The following assumptions for the conversion of  coefficients and carbon units were used: we 
\begin{inparaenum}[(i)] 
\item take the experimentally-determined organic matter fraction of $56\%$ (average over all experiments
in \citealt{Bayne1993} of measured particulate organic matter (POM) to \TPM);
\item use carbon mass to molar ratio with of 12.011\,mg mmol$^{-1}$;
\item use dry weight (DW), ash free dry weight (AFDW) and wet weight conversions from \citet{Ricciardi1998};
\item apply molar mass
conversion in Redfield stoichiometry (molar ratio 106:16:1 C:N:P) to express the DW
to amount carbon ratio as $32.43$\,mg
per mmol\,C;
\item scale all rates to individual mass 600\,mg with the experimentally 
confirmed metabolic scaling exponent of $0.67$ \citep{Bayne1980,Bayne1993}.  
\end{inparaenum}
As a lower threshold for filtration, a
phytoplankton carbon concentration of $[\mathrm{C}]_{\min}=0.7$\,\mpcbm{} was chosen, 
consistent with the threshold suggested by  \citet{Riisgard2003} of  $0.5$\,mg Chl-a m$^{-3}$.
Filtration of phytoplankton biomass by \edulis{} removes particulate carbon, nitrogen, and phosphorus from the pelagic  phytoplankton compartment, in the  same stoichiometric proportion as the food, and with it also reduces  dependent phytoplankton properties like Chl-a.  The phytoplankton compartment is
converted to detritus, representing faeces and pseudofaeces, in carbon, nitrogen and phosphorous.  We assume
that 20\% of the carbon is lost to respiration, leading to higher quality ejected detritus compared to the food source; direct DIN (e.g., urea, see \citealt[e.g.,][]{Cockcroft1990}) release by mussels is not considered. 

The filtration model is technically realised as an Earth System Modeling Framework \citep[ESMF,][]{Hill2004} component 
and coupled with MOSSCO \citep{Lemmen2018} to the Framework for Aquatic Biogeochemical Models 
\citep[FABM,][]{Bruggeman2014} with the MAECS biogeochemical model in the pelagic and
OmexDia \citep{Soetaert1996a} with added
phosphorous cycle \citep{Hofmeister2014iche} in the soil FABM domains. 

\subsubsection{Model setup and scenarios}

The SNS was represented on a curvilinear grid
with cell size between $2$ and $64$\,km$^{2}$, with the highest resolution in the
German Bight. Vertically, the water column was
represented by 20~terrain-following $\sigma$-layers \citep{Kerimoglu2017}.
The model setup  accounts for the discharge of freshwater, phosphorous and nitrogen from
major rivers into the southern North Sea, including the Elbe, Weser, Ems, Rhine, Meuse,
Scheldt and Humber \citep[see][]{Kerimoglu2017}. Tidal sea surface
elevation was forced at the open ocean boundary.  Open ocean boundary conditions for
nutrients in dissolved and particulate forms were obtained from a North Atlantic shelf
simulation with  ECOHAM (Ecosystem Model Hamburg, \citealt{Grosse2016}) and
provided as a 10~year climatology \citet{Kerimoglu2017}. Phytoplankton and zooplankton
were assumed to be at zero-gradient at the boundaries.
The meteorological forcing was obtained from the long-term Climate Limited area
Model reconstruction available in the CoastDat database \citep{Geyer2014}.

\begin{table}
\small\centering
\caption{Scenarios contrasted in this study}
\begin{tabular}{lp{.5\hsize}p{.2\hsize}}\hline
Scenario & Description & Total biomass\\ \hline
REF & Presence of epibenthic mussels.  This represents the reference
state against which the addition of artificial hard substrate by OWFs is compared. & \science{16}{11}~indiv.  96~tons \\
OWF  & As REF, but with additional presence of epistructural mussels in pelagic surface layers. & \science{7}{10}~indiv.  42~tons\\ \hline
\end{tabular}
\label{tab:scenarios}
\end{table}

Simulations were run for the duration for 14 consecutive years 2000-2013, with the first three years
discarded to allow for a model spin-up, especially for the equilibration of winter nutrient storage in 
the sediment.  As we are evaluating a sensitivity for a projected year 2030 scenario, the choice of
this period is arbitrary and reflects availability of station and satellite data for model evaluation. Two
different scenarios were compared, 
\begin{inparaenum}[(1)]
\item presence of only epibenthic mussels (scenario ``REF''), and 
\item additional presence of epistructural \edulis{} at OWFs, focussed within the upper
pelagic layers (scenario ``OWF'')
\end{inparaenum} (\reftab{scenarios}).

The filtration model was configured with phytoplankton carbon as the main species to filter,
with co-filtration of phytoplankton nitrogen, phosphorous, Chl-a and rubisco. 
The model diagnostic rates of relative carbon uptake were multiplied by phytoplankton
carbon concentration and subsequently integrated for the entire year to obtain the annual
net primary productivity. 
The  3D time step of the hydrodynamic model was 6~minutes.
Data exchange between the different components of the model system was performed 
every 30~minutes. The bottom roughness length was constant at
$z_{0}=0.002\,\mathrm{m}$, wave forcing was disabled.  A Jerlov Type~III water class
was used for the radiation scheme. 

\subsection{Data for model evaluation}
No observational data is available for primary productivity at the scale of the SNS. Rather than
productivity as a rate, the stock of phytoplankton is readily observed with \emph{in situ} methods 
or by remote sensing.  We evaluate Chl-a as simulated  by the model against
station observations of chlorophyll fluorescence along three transects and against 
synoptic satellite observations of ocean color. 

Time series of near-surface Chl-a concentration were obtained from the Dutch authority
Rijkswaterstaat through the OpenEarth portal \citep{Rijkswaterstaat2017}. From all available
station data, we selected three transects that cross the coastal nutrient gradient from nearshore Noordwijk, 
Terschelling and Rottumerplat to up to 235\,km offshore.  Satellite observations
were obtained from the European Space Agency Ocean Color Climate Change Initiative (ESA-CCI version~3.1), a 
multi-platform combined product of Chl-a concentration.  

\section{Results}\label{sec:results}

\begin{figure*}
\begin{overpic}[viewport=0 90 430 590,clip=,width=\textwidth]{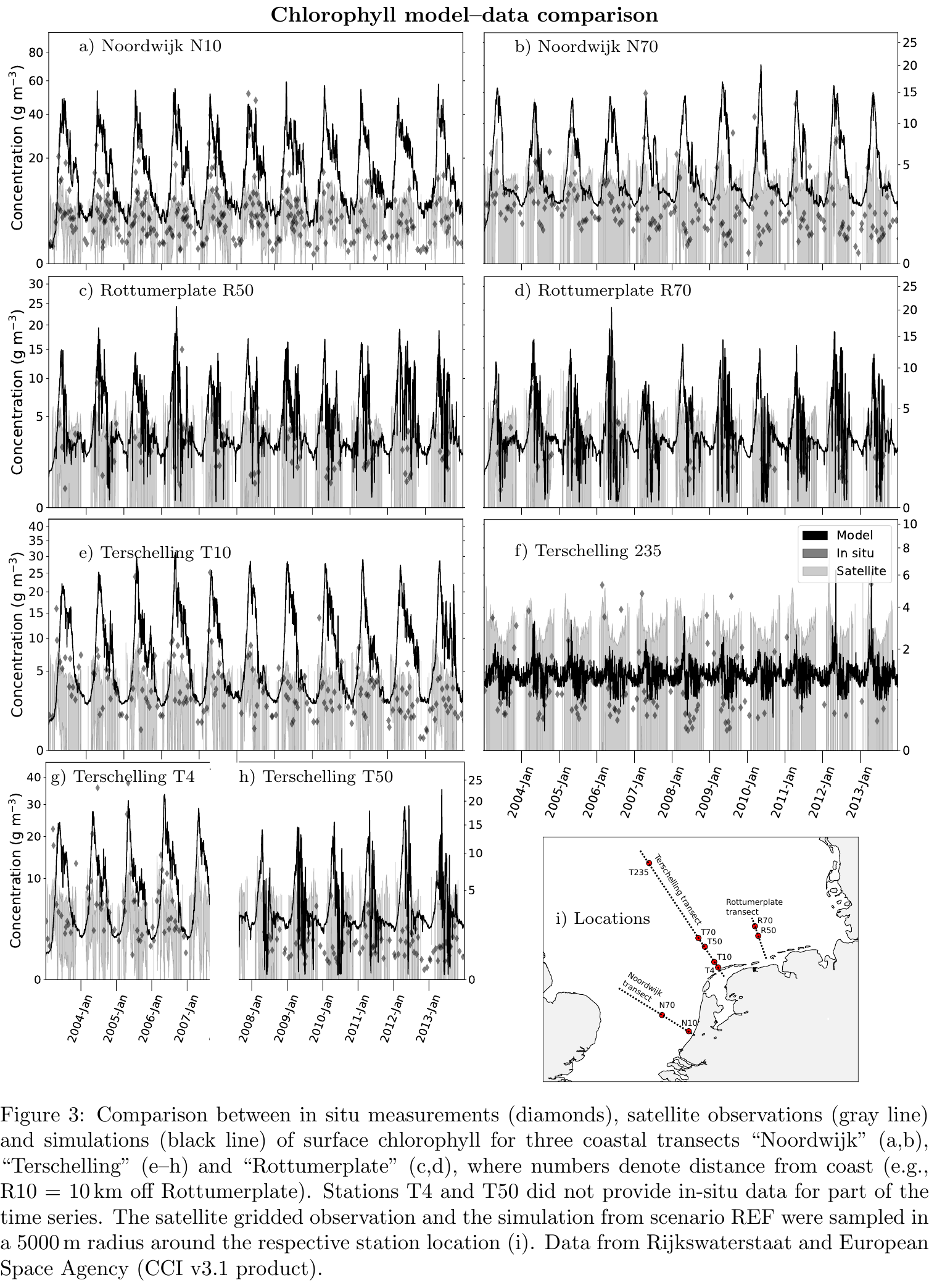}
\end{overpic}
\caption{Comparison between in situ measurements (diamonds), satellite observations (gray line) and simulations (black line) of surface chlorophyll for three coastal transects ``Noordwijk'' (a,b), ``Terschelling'' (e--h) and ``Rottumerplate'' (c,d), where numbers denote distance from coast (e.g., R10 = 10\,km off Rottumerplate). Stations T4 and T50 did not provide in-situ data for part of the time series.  The satellite gridded observation and the simulation from scenario REF were sampled in a 5000\,m radius around the respective station location (i).  Data from Rijkswaterstaat and European Space Agency (CCI~v3.1 product).}
\label{fig:stations}
\end{figure*}

The reconstructed abundance of \edulis{} in the SNS suggests
$1.6\cdot 10^{11}$ individuals on natural (benthic) substrate and within mussel beds (\reffig{musseldistribution}b)
The reconstructed accumulated biomass of benthic 
\edulis{} in the SNS amounts to  total mussel mass of 96 Mt DW  (or  10 Mt AFDW).  
For the potential ``artificial'' stock at offshore wind turbines,
the reconstructed abundance (\reffig{musseldistribution}c)  in the entire SNS amounts
to \science{7.0}{10} individuals, or 42 Mt DW (4.5 Mt AFDW).  
Once all the planned wind farms are in operation,
they will provide habitat for mussels that are equal to 44\% of the  stock of
benthic mussels.

\subsection{Uncertainty estimates of reconstruction}
The reconstruction of mussel abundance in the southern North Sea is based on
analysis of field data (in total 4074 count observations and 37\,214
presence only data, which reveals a positive
correlation ($r=0.78$) between abundance and sediment grain size. The 10\,m
water depth line is introduced to provide a
pseudo-absence criterion.  To test a sensitivity of the reconstruction result to the water depth limitation,
we also calculated the abundance using 25\,m water depth
contour line ($\approx$95\% of observed presence occurs within this water depth)
as an alternative constraint, which leads to an increase in abundance by 
$\approx\science{1.4}{10}$ compared to that using the 10\,m. This amounts to $\approx$9\% of the
total budget estimated using the 10\,m water depth constraint, and thus does not
affect  our estimation to a large degree. The area covered by mussel beds in the Wadden Sea
oscillated annually between \science{6.2}{7} and \science{3}{7}\,m$^{2}$ since
1998 \citep{Nehls2009}. Since our estimation of the stock in the Wadden Sea
is based on a maximum value, the annual oscillation
of mussel beds would affect the total estimated budget in the SNS by 10\% at 
most: our reconstruction and the annual fluctuation should be within 30\% of the total
estimated budget, taking into account oscillations of the mussel beds and the
impact of extreme wind waves on offshore wind turbines, which might 
occasionally clear all mussels from a wind turbine \citep{Krone2013}.

\subsection{Simulated chlorophyll and comparison to station/statellite data}
Simulated surface Chl-a for the years 2003--2013 exhibits a typical annual
phytoplankton cycle with a large spring bloom and a smaller summer bloom. At
the bloom peak, the Chl-a concentration reaches 20--50\,mg \,m$^{-3}$ at coastal
and below 5\,mg\,m$^{-3}$ at far offshore locations (\reffig{stations}).  The
simulation reproduces \emph{in situ} time series of near-surface
Chl-a concentration along the three transects from Noordwijk, Terschelling and
Rottumerplate. The peak spring bloom Chl-a concentrations are well matched across
the entire coastal gradient; overall the simulation has a small positive bias below 4\,mg \,m$^{-3}$, with 
a larger overestimation of 9\,mg \,m$^{-3}$  at Noordwijk 10.
The variability of Chl-a concentrations is also well represented. At most stations, the simulated and
the observed standard deviation agree to within 1.2\,mg \,m$^{-3}$, with the exception of Terschelling 10 and
Noordwijk 10, where the model standard deviation is 3\,mg \,m$^{-3}$ higher than the observed variability.

The comparison against satellite observations shows that both model and \emph{in situ} observations
have a wider temporal variability, while the mean Chl-a concentration is again well represented.
The model surface Chl-a climatology, i.e. the multi-annual average over all years 2003--2013, 
has a small positive bias compared to
satellite observations: it is below or near $1.0$\,mg \,m$^{-3}$ in fall and winter, and largest during May, when 
simulated Chl-a is $3.6$ \,mg \,m$^{-3}$ larger on average. This difference is smallest (below $1$\,mg \,m$^{-3}$)
offshore and where most of the OWF are located. It is largest (up to  $15$ \,mg \,m$^{-3}$) in the
near-shore high-productivity zone along the East and West Frisian barrier islands. 

\subsection{Net primary productivity}

\begin{figure*}
\centering
\begin{overpic}[viewport=0 30 430 320,clip=,width=\hsize]{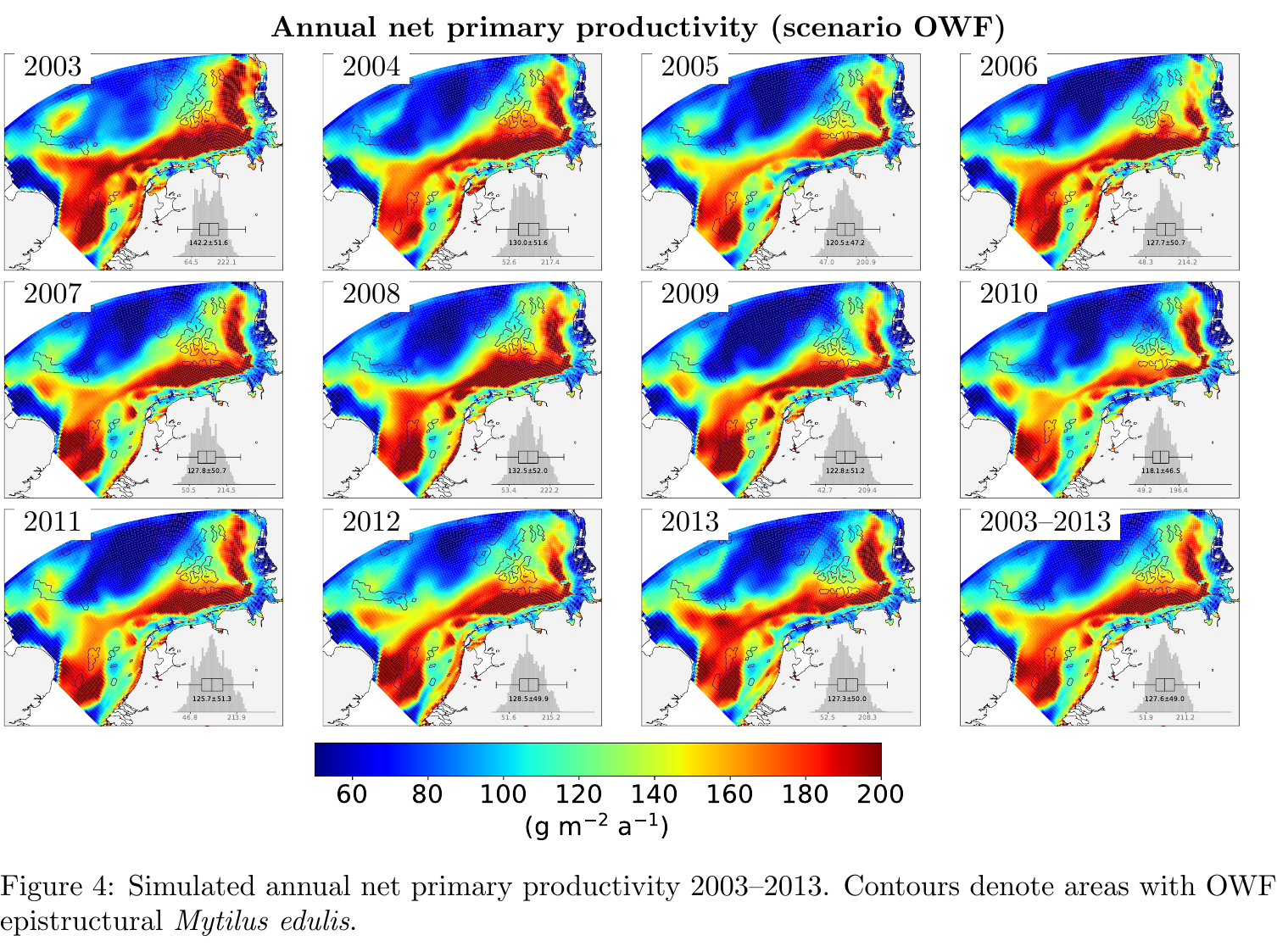}
\end{overpic}
\caption{Simulated annual net primary productivity 2003--2013.  Contours denote areas with OWF epistructural \edulis{}.}
\label{fig:nppclimatology}
\end{figure*}

The simulated annual vertically integrated net primary productivity (NPP, expressed as carbon
productivity) in the
SNS, as well as the climatological average over the years 2003--2013, broadly separates the
model domain into three 
regions (\reffig{nppclimatology}): 
\begin{inparaenum}[(1)]
\item the coastal area including the Wadden Sea, of highly variable and low vertically integrated
carbon production ($<50\,\gpy$, very shallow and turbid water), 
\item the near-coast transition zone
with a high productivity above 180\,\gpy{}  up to  $\approx400\,\gpy$, and
\item the offshore SNS,
again with relatively low productivity around $90\,\gpy$ .  
\end{inparaenum}

This  pattern is consistent across all simulation years.  Maximum productivity in this simulation occurs
in an elongated coast-following area 20\,km
north and east of the West Frisian and East Frisian islands, in the central Southern Bight, and off the East
coast of England.  Within the period 2003--2013, the year 2010 exhibits the
lowest productivity with $118\pm47\,\gpy$, and the it is highest in 2003 ($142\pm52\,\gpy$).  Most OWF 
are located in the transition zone between the maximum productivity band and the low productivity
areas offshore.

\begin{figure*}
\centering
\begin{overpic}[viewport=0 45 435 340,clip=,width=\hsize]{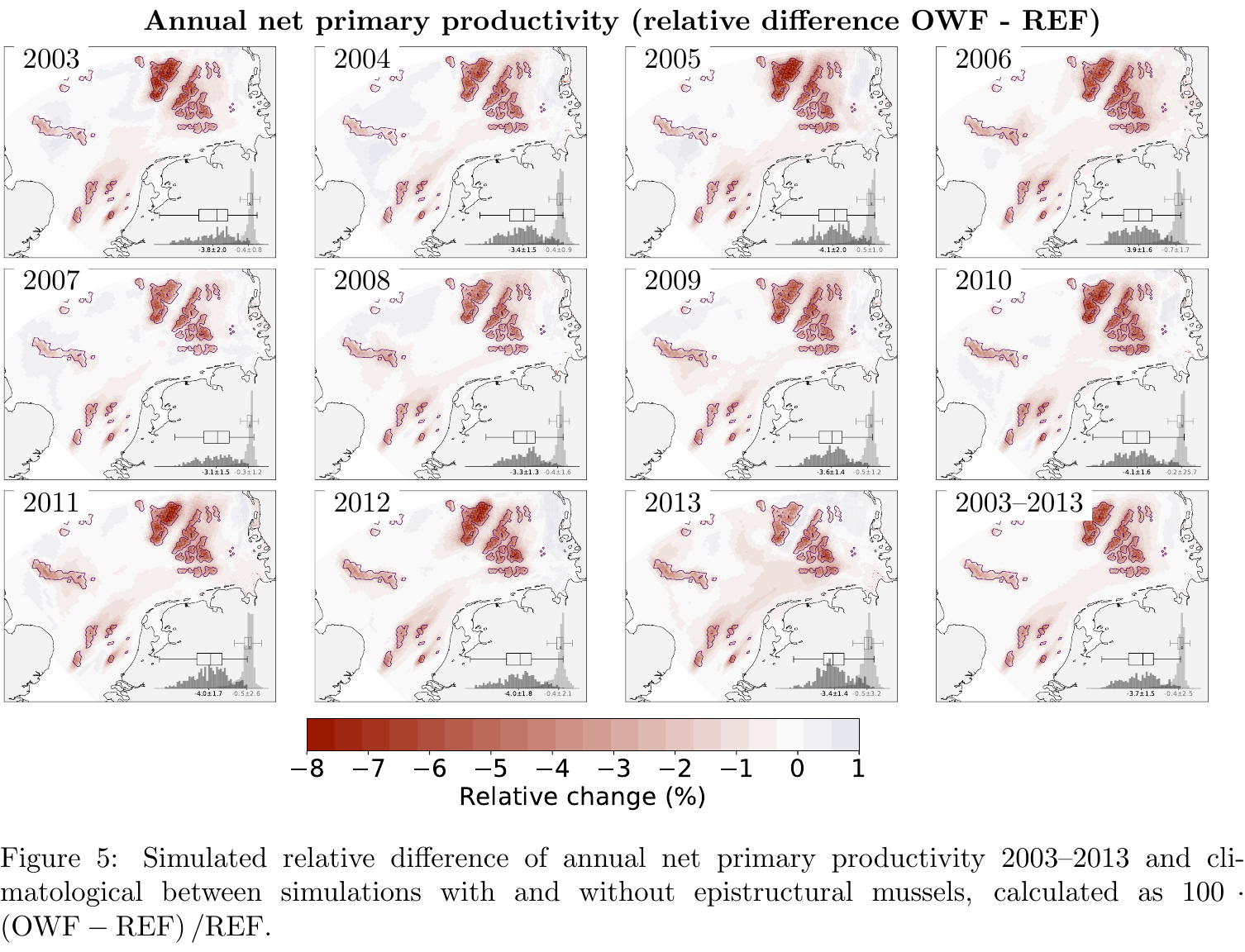}
\end{overpic}
\caption{Simulated relative difference of annual net primary productivity 2003--2013 and climatological between simulations with and without epistructural mussels, calculated as $100 \cdot \left(\textrm{OWF} - \textrm{REF}\right) / \textrm{REF}$.}
\label{fig:nppdiff}
\end{figure*}

There is less primary productivity locally in the OWF than in the REF scenario in all years
(\reffig{nppdiff}).  The maximum loss occurs within the OWF areas (up to $8\%$), and is on average 
$3.7\pm1.5\%$, with a maximum in 2005 and 2010 ($4.1\%$) and a minimum in 2008 ($3.3\%$).  Variability
is high between the different OWF areas  with a climatological standard deviations of $1.5\%$).  Loss
outside the OWF areas is much smaller, but the change is consistently negative and $0.4\pm2.5\%$) in the long-term
mean.  This outside-OWF loss also has a typical distribution with largest losses in the maximum productivity band
along the East and West Frisian barrier islands and in the vicinity of the OWF.  In many years, productivity is
increased (a very small increase below $1\%$) along the North Frisian barrier islands. 

\begin{figure}
\centering
\label{fig:phyc}
\begin{overpic}[viewport=0 50 435 530,clip=,width=.75\textwidth]{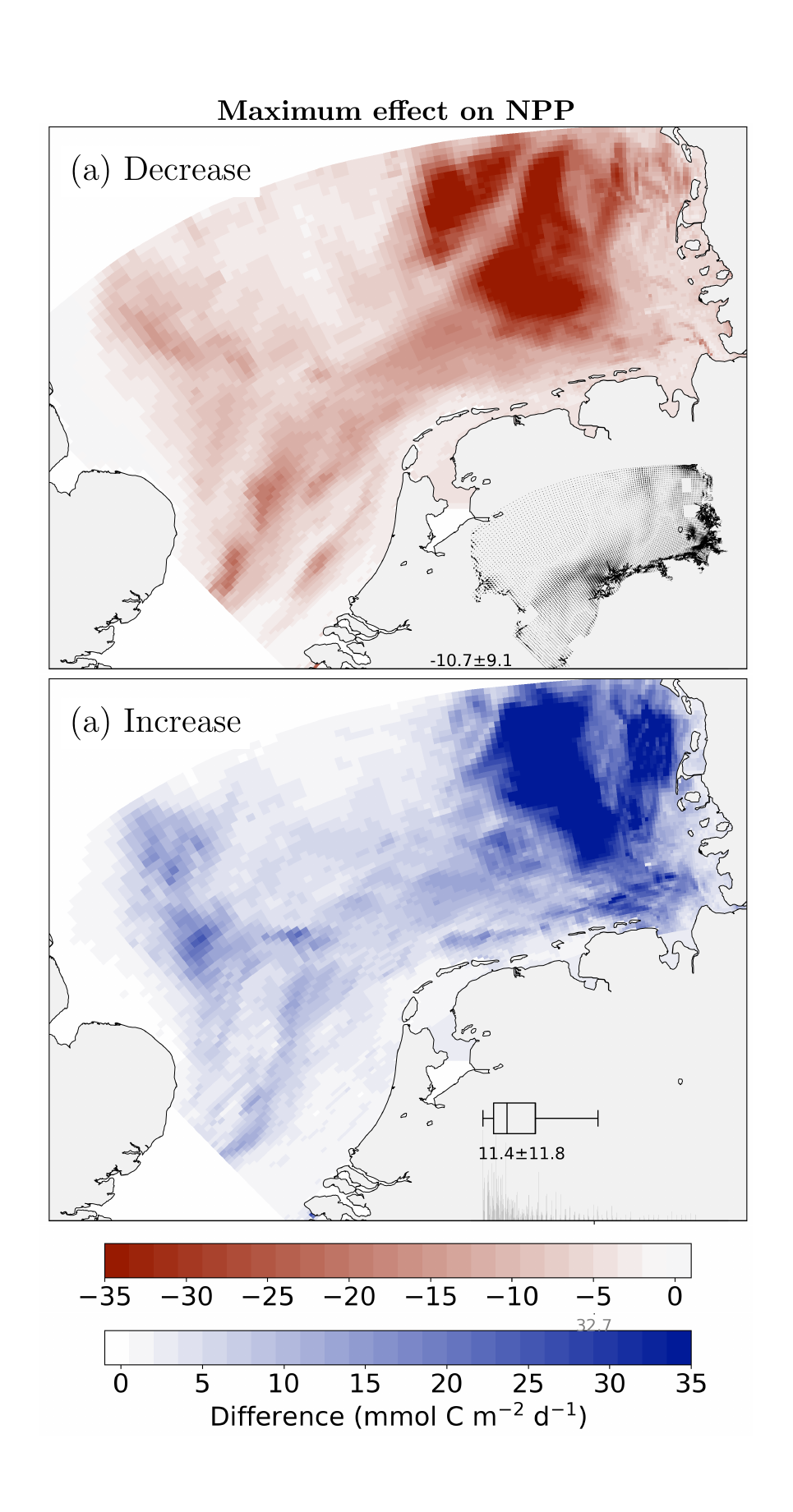}
\end{overpic}
\caption{Maximum daily net primary productivity effect of epistructural \edulis{} (exemplary for year 2006). (a)  Maximum decrease in 2006, and (b) maximum increase in 2006.}
\end{figure}

To identify a regional effect outside the OWFs, we identified the maximum increase and maximum decrease of daily NPP between the scenarios for each year (\reffig{phyc}, shown for 2006).  The maximum daily decrease of  NPP is $-11 \pm 9$\,\mpcbmd, with the largest decreases (below $-20$\,\mpcbmd)  occurring within the two large clusters of OWF areas in the eastern SNS.  The spatial distribution of the maximum daily increase of NPP shows changes of the same order of magnitude throughout the SNS ($11 \pm 12$\,\mpcbmd).  In contrast, however, maximum increases also occur outside the OWF areas, with the largest increases (above $20$\,\mpcbmd) east of the central eastern SNS OWF cluster and also bear the East Frisian and North Frisian barrier islands.  The maximum increase is seen up to  50\,km away from the farms.

\section{Discussion}\label{sec:discussion}

Net primary productivity calculated by our coupled model shows low productivity in the Wadden
Sea area (\reffig{nppclimatology}). Simulated Chl-a concentrations in this area are also
lower than those estimated by the satellite imagery \citep[e.g.,][]{Kerimoglu2017,Ford2017},
while in the open SNS, our model modestly overestimates Chl-a and  probably also NPP. 
NPP simulated by \citet{VanLeeuwen2013} with the same hydrodynamic (GETM) but with a
different ecosystem model (ERSEM, \citealt{Baretta1995}) is  much higher (on average
$318\pm 29$\,\gpy) than NPP simulated here for their region termed ``SNS'', referring to a small area
of the Southern Bight of the North Sea.  While this is not a good choice of region for comparison, because the exact location of this maximum productivity region varies between the years (\reffig{nppclimatology}), also their entire North Sea estimate of $180\pm 10$\,\gpy{}   is higher than our calculation.  When comparing the two studies, however, one should note that they averaged
over the much higher trophic state period 1985--2005, such that lower productivity should be
expected for the period 2003--2013. 

Already  \citet{Emeis2015} report values around 200--270\,\gpy, for an area corresponding to our
coastal high productivity region in the year 2002, based on the Ecosystem Model
Hamburg (ECOHAM, \citealt{Paetsch2008}).  The comparison by \citet{VanLeeuwen2013}
with \emph{in situ} observation derived NPP estimates by \citet{Weston2005},
however, also showed that their model under- or overestimates observations by a factor of two
depending on the area type (stratified, frontal bank), and overestimated surface mixed layer
productivity by up to a factor of five \citep[][Table 1]{VanLeeuwen2013}.  
Given these considerations the simulated productivity in the
coastal and open SNS (ranging between 50 and 400\,\gpy{}) is plausible.  Its skill needs to
be assessed against observational data in forthcoming studies, such as Wirtz (submitted) .

In contrast to productivity, biomass related variables are readily observable from a variety of platforms:
the agreement between the \emph{in situ} measured, the remote sensing observed, and 
the simulated Chl-a
concentrations (\reffig{stations}) suggests that phytoplankton dynamics is well
reproduced, which builds confidence in the representation of primary productivity
by the model.
Moderate discrepancies in the cross-coastal distribution of NPP and Chl-a are in part due
to the simplified description of water attenuation by high concentrations of suspended
particles  close to the coast \citep{Kerimoglu2017} and the prescribed mortality gradient of zooplankton. 
However, given that the concentrations are in the range where mussel
filtration exhibits a linear functional response to food concentration, the moderate mismatch should not affect  our estimates of the relative effects of additional epistructural filtration. 

\subsection{Non-local spatial effects}\label{sec:trends}
During a bloom, phytoplankton will usually first appear at the surface
and then sink down through the water column \citep{Cloern1996}, producing a delay
between surface phytoplankton concentrations and those at depths.  Similarly, the
phytoplankton loss from epistructural filtration is first and clearly visible  at the surface,
where \edulis{} are concentrated, before being transmitted into the entire water column. 
After filtration, nutrients that were bound in phytoplankton are readily made available by
pelagic remineralization of the ejected high-quality detritus.  By this mechanism, it is to 
be expected that filtration sustains a longer bloom through faster nutrient recycling and also
supports higher productivity in regions that receive nutrient-enriched and phytoplankton reduced
water masses from  OWF areas by currents.  

The maximum daily NPP changes (\reffig{phyc}) indeed demonstrate that the ecosystem effect of epistructural filtration is not a local one, but a regional one, with a decrease of phytoplankton carbon throughout many parts of the SNS (albeit concentrated up to 20\,km around the OWF) and
a strong increase up to 50\,km outside the OWF area.  
It can be argued that the magnitude of several percent per year in
overall draw-down  is well within the uncertainty range of state-of-the-art ecosystem models.  The effect is, however, regionally very different and thus changes horizontal gradients in productivity that have not been discussed before: there is a notable impact of projected epistructural suspension feeders on the ecosystem 
functions of a regional shelf sea. Even though the decrease in primary productivity is relatively small, it
extends over a large area and intensifies in close proximity to OWFs, reaching a maximum reduction in
annual net primary productivity of 8\%.  Despite the dilution of meso-scale mussel density due to the
low area density of offshore wind turbines, massive biofouling accumulates to an effect size which
is only one order of magnitude lower than the 60\% reduction within shellfish aquacultures \citep{Waite1989}. 

\subsection{Altered ecological functioning}\label{sec:ecologicalfunctioning}

Primary productivity represents the major energy source for ecosystems globally \citep{Imhoff2004}.  Our
model results indicate that the construction of OWFs reduces available primary productivity,
especially at the local scale, as a result of a higher abundance of filter feeders \citep{Prins1997}.  
Filtration activity transforms the carbon, nutrient and energy flows at OWFs from which the benthic food-web
benefits, with faeces, pseudofaeces and dead mussels enriching the
benthic organic pool, as observed in many shellfish aquaculture facilities worldwide \citep[e.g.][]{Cranford2007}.
Notably, filter feeding much accelerates the transformation of living biomass into partially dissolved and bio-available nutrients, which may again fuel productivity. The effects on autotrophic standing stock investigated in our study hence do not provide a full account of processes relevant for assessing primary productivity. 

Our results suggest that the increased abundance of \edulis{} at OWFs only moderately affects ecosystem
functioning. They emphasize \edulis{}'s role as an ecosystem engineer \citep{Borthagaray2007}, not just
locally but on a scale of greater than 50~km. Pelagic primary productivity is just one of many facets of ecosystem
functioning, which have been hypothesized to be altered by OWFs (amongst others):
\begin{asparaenum}
\item Networks of OWFs are beneficial for the conservation of threatened species by acting as \emph{de facto}
marine protected areas \citep{Inger2009}.  Access to most areas designated as OWFs is limited, primarily for safety
reasons, which limits anthropogenic pressures such as fishing, including bottom trawling, potentially leading to an
increased level of biodiversity at OWFs as compared to unprotected areas  \citep{Kaiser2006,Tillin2006,Inger2009}.  
\item Mussels such as  \edulis{} play a significant role in modifying the natural substrate and increasing species
richness \citep{Borthagaray2007}.  \edulis{} bioengineers its surrounding environment through shell litter fall 
\citep{White1985}, water filtration and clarification  \citep{Newell2004a}, concentration of nutrients
\citep{VanBroekhoven2014}, ultimately increasing the species diversity and landscape heterogeneity as a result of
structural and functional effects  \citep{Norling2007}. Shells, both dead or living, increase the degree of habitat
complexity, encouraging a higher level of species richness  \citep{White1985}.  Bivalve and gastropod shells are
persistent and abundant physical structures which provide substrata for attachment and refuge from predation as
well as physical or physiological stress  \citep{Gutierrez2003}.  
\item Mediated through the associated epibenthic community, OWF constructions can
act as stepping stones for the dispersal of exotic species  \citep{Glasby2007}.  The artificial habitat is open for
colonization by new species assemblages \citep{Wilson2009}, which could not quickly establish in soft-bottom
sea regions.  One example of this is \species{Telmatogeton japonicus}, the marine splash midge, which is native
to Australasian waters.  Known to be transported on the hulls of ships, it has been observed at OWFs in
Denmark and along the Swedish Baltic coast \citep{Wilhelmsson2008}.  The projected density of offshore
constructions will likely facilitate immigration by  non-native species  \citep{Bulleri2005}, such as the leathery
sea squirt (\species{Styela clava}), slipper limpet (\species{Crepidula fornicate}), pacific oyster
(\species{Crassostrea gigas}) and Japanese skeleton shrimp (\species{Caprella mutica}) 
\citep{Buschbaum2005,Diederich2005,Luetzen1999,Thieltges2003}.  Through these changes in biodiversity,
OWFs could shape the marine ecosystem beyond their physical boundaries.
\item \edulis{} are a highly diverse prey source capable of supporting higher trophic levels, especially
vagile demersal megafauna (e.g. fish like \species{Trisopterus luscus} and crabs like \species{Cancer pagurus}) 
\citep{Langhamer2009,Reubens2011}.  Their abundance and distribution at OWFs is driven by changes in attraction,
productivity and redistribution  \citep{Bohnsack1989}.  An increase in the population of vagile demersal megafauna
further impact the local community, leading to increased species diversity  \citep{Wilhelmsson2008}.  Larger
megafauna may also benefit from increased food availability  \citep[e.g.,][]{Brasseur2012}, with seals extending their
distribution towards an OWF after construction in Denmark  \citep{Tougaard2006}.
\end{asparaenum}
 
Many of the ecosystem feedbacks and hence changes to ecosystem services
 are yet unknown and need to be studied both \emph{in situ} and in future
system-wide synoptic studies.  Mostly for supporting the planning process, a high number of often not published
studies were conducted, almost always considering individual offshore wind turbines and focused on
selected potential effects such as on birds, megafauna, or hydrodynamics \citep[e.g.]{Bailey2014}. 

\subsection{Limitations and outlook}\label{sec:}

This study is the first large scale assessment of epistructural \edulis{} filtration at OWFs.  
The level of quantification achieved in our study clearly shows that at least with respect
to primary productivity there is a non-negligible regional ecosystem effect originating from epistructural
\edulis{} filtration.   Modular model systems are needed to integrate effects and feed-backs across trophic levels
and up to the regional scale, as proposed and to a large but not complete degree realised here.

There are still large uncertainties related to simulating  complex ecosystem interactions.   The reconstruction of filter
feeder epistructural abundance is yet a simple up-scaling of data sampled from single piles. And the 
benthic reconstruction suffers from the sampling bias introduced by opportunistic observations.  These data 
issues will likely ameliorate in the future with monitoring programs and systematic surveys.  The filtration 
model   is very simple and does not include, for example, age structured population dynamics or
nutrient recycling: this study shows how essential it is to  improve filtration models, which so far are often neglected
in shelf ecosystem models. For  studies adressing the forthcoming decades, a more
accurate  quantification of the epistructural filtration effect seems to be required.  Physical effects of wind farm on
atmospheric boundary
layer circulation and ocean currents and vertical mixing  \citep[e.g.][]{McCombs2014,Carpenter2016} have not been
considered by our coupled model: there is still a scale problem
that needs to be addressed in physical modelling to bridge the wind pile (order of 10 m) to ecosystem (order of
100 km) scales. Recent developments in nested or unstructured models seem promising here. Last but not least,
the uncertainties of the simulation result are difficult to quantify:   estimates of productivity in the literature have
large uncertainties themselves. There are no regional studies to which the filtration rates can be compared.  
Our findings of a 8\% annual productivity and up to 30\mpcbmd daily productivity 
justify further research into the large-scale impact of OWFs. Remote sensing observations might provide
the first regional direct observations of OWF effects as the resolution of the sensors now allows the
identification of single wind turbines, as has already been done for physical parameters \citep{Platis2018}. 

\section{Conclusions}\label{sec:conclusion}

Construction of offshore wind farms (OWFs) in the southern North Sea is predicted to continue into the future,
highlighting the importance of understanding the potential nature and magnitude of the environmental impact
of the epistructural biomass known to accumulate on their subsurface structures.  Epistructural communities at OWFs
in the southern North Sea are dominated by \edulis{}, a filter feeder capable of inducing extensive ecological
change through filtration, amongst other processes.  The construction of all currently planned, consented and
under construction OWFs, in addition to those already in operation, is likely to increase the overall abundance
of \edulis{} in the southern North Sea by more than $40\%$.  In addition to providing an additional food source
and a new habitat, \edulis{} also remove phytoplankton from the water column through filtration, which
impacts ecosystem functioning.

Reconstructing and analysing the impact of epistructural biomass at OWFs on pelagic primary productivity at a larger
spatial scale, in this case the southern North Sea, provides valuable insights into ecosystem functioning which
are not visible at the scale of a single offshore wind turbine or OWF.  The impact of
OWFs on annual primary productivity is predominately local, at short time scales there is a larger regional effect on
biomass and productivity that extends up to several 100s of km beyond the bounds of the OWF area.  

\section*{Code and data availability}

{Development code and documentation are hosted on Sourceforge
(\href{https://sf.net/p/mossco/code}{https://sf.net\-/p/mossco\-/code})
The release version 1.0.3 is permanently archived and accessible under the digital object identifier\\
\href{https://doi.org/10.5281/zenodo.1243045}{https://doi.org/10.5281/zenodo.1243045}.  All external software used
is available as open source and can be obtained together with MOSSCO.  The simulations were
performed in parallel on 192~cores on the Jureca high performance computer at Forschungszentrum
J\"ulich, Germany \citep{Krause2016}.}

Satellite data are freely available from the Ocean Colour Climate Change Initiative dataset, Version 3.1,
European Space Agency,  at \href{http://www.esa-oceancolour-cci.org/}{http://www.esa-oceancolour-cci.org/}. 
Meteorological forcing data are available on request from CoastDat at \href{http://www.coastdat.de}{www.coastdat.de} \citep{Geyer2014}. 
Chlorophyll a and other water quality parameters are freely available at \href{http://www.waterbase.nl}{http://www.waterbase.nl} \citep{Rijkswaterstaat2017}.

The reconstructed epibenthic and epistructural \edulis{} maps (\reffig{musseldistribution}) and the simulated net primary productivity data for both scenarios (Figs.~\ref{fig:nppclimatology}, \ref{fig:nppdiff}) have been archived with PANGAEA --- Data Publisher for Earth \& Environmental Science as
the dataset ``Simulated net primary productivity (NPP) in the southern North Sea 2003-2013 forced by epistructural and epibenthic reconstructed blue mussel filtration''.  They are available under the digital object identifier\\ \href{https://doi.org/XXXXXX.XXXX}{https://doi.org/XXXXXX.XXXX}.


\bibliographystyle{spbasic_hydr}
\bibliography{Slavik2018_etal_hydrobiologia}   

\end{document}